\documentclass[10pt,journal,twocolumn,twoside]{IEEEtran}
\usepackage{optional}

\usepackage[cmex10]{amsmath}
\usepackage{amssymb}

\usepackage{array}
\usepackage[caption=false,font=footnotesize]{subfig}
\usepackage{fixltx2e}
\usepackage{url} 

\usepackage{graphicx}
\usepackage{algorithmic}
\usepackage{rotating}
\usepackage{multirow}
\usepackage{hyperref}
\usepackage{epstopdf}
\usepackage[noadjust]{cite}

\renewcommand{\vec}{\mathbf}
\newcommand{\greekvec}{\boldsymbol}
\begin{document}
  \title{Sequential Complexity as a Descriptor for \\ Musical Similarity}
  \author{
      Peter~Foster,~\IEEEmembership{Student~Member,~IEEE,}
      Matthias~Mauch,~\IEEEmembership{Member,~IEEE}
      and~Simon~Dixon%
	  \thanks{Copyright~\copyright~2014 IEEE. Personal use of this material is permitted.
	  However, permission to use this material for any other purposes must be
	  obtained from the IEEE by sending a request to pubs-permissions@ieee.org.}
      \thanks{P.F.~is funded by an Engineering and Physical Sciences Research Council Doctoral Training Account studentship. M.M.~is funded by a Royal Academy of Engineering Research Fellowship.}
      \thanks{All authors are with the School of Electronic Engineering and Computer Science, Queen Mary University of London, London, E1~4NS, UK. (Email: peter.foster@eecs.qmul.ac.uk; matthias.mauch@eecs.qmul.ac.uk; simon.dixon@eecs.qmul.ac.uk)}}

\markboth{IEEE/ACM Transactions on Audio, Speech and Language Processing, VOL.~22, NO.~12, December 2014}%
{Foster \MakeLowercase{\textit{et al.}}: Sequential Complexity as a Descriptor for Musical Similarity}

 \maketitle

\begin{abstract}
We propose string compressibility as a descriptor of temporal structure in audio, for the purpose of determining musical similarity. Our descriptors are based on computing track-wise compression rates of quantised audio features, using multiple temporal resolutions and quantisation granularities. To verify that our descriptors capture musically relevant information, we incorporate our descriptors into similarity rating prediction and song year prediction tasks. We base our evaluation on a dataset of 15\,500 track excerpts of Western popular music, for which we obtain 7\,800 web-sourced pairwise similarity ratings. To assess the agreement among similarity ratings, we perform an evaluation under controlled conditions, obtaining a rank correlation of 0.33 between intersected sets of ratings. Combined with bag-of-features descriptors, we obtain performance gains of 31.1\% and 10.9\% for similarity rating prediction and song year prediction. For both tasks, analysis of selected descriptors reveals that representing features at multiple time scales benefits prediction accuracy.
\end{abstract}

  \begin{IEEEkeywords}
  Music content analysis, musical similarity measures, time series complexity
  \end{IEEEkeywords}

  \ifCLASSOPTIONpeerreview
  \begin{center} \bfseries EDICS Category: AUD-CONT \end{center}
  \fi


\section{Introduction} 
We are concerned with the task of quantifying musical similarity, which has received considerable interest in the field of audio-based music content analysis \cite{casey2008content,fu2011survey}. Owing to the proliferation of music in digital formats and the expansion of web-based music databases, there is an impetus to develop novel search, navigation and recommendation systems. Music content analysis has found application in such information retrieval systems as an alternative to manual annotation processes, when the latter are infeasible, unavailable or amenable to be supplemented \cite{celma2009music}.

We may distinguish between music content analysis applications such as audio fingerprinting \cite{cano2005review}, version identification \cite{serra2011identification}, genre classification \cite{scaringella2006automatic} and mood identification \cite{kim2010music}. Given a query track, audio fingerprinting typically should identify a unique track deemed similar with respect to a collection. In contrast, for genre and mood classification, the set of tracks deemed similar with respect to a collection is typically large. Thus, we may distinguish between music classification tasks according to the degree of \textit{specificity} associated with the measure of musical similarity \cite{casey2008content}.

In this work, we consider two low-specificity tasks, namely similarity rating prediction and song year prediction. An important issue in our considered domain surrounds feature representation. In particular, we address the problem of representing temporal structure in audio features. We refer to summary statistics of audio features extracted from a song as descriptors. Descriptors may be characterised according to how temporal structure is accounted for \cite{fu2011survey}. We may distinguish between \textit{bag-of-features} representations \cite{aucouturier2007bag}, which discard information on temporal structure, and sequential representations. As a sequential representation, we propose to estimate the complexity of audio feature time series, where we quantify complexity in terms of string compressibility. As a result, we obtain scalar-valued summary statistics which retain information on temporal structure.

We motivate our evaluations involving similarity rating prediction and song year prediction to test the hypothesis that our complexity descriptors capture temporal information in audio features and that such information is relevant for determining musical similarity. For similarity rating prediction, our ground truth is given by human similarity judgements and we assume that an objective musical similarity correlates with subjects' degree of perceived musical similarity, based on a five-point rating scale. For song year prediction, our ground truth is readily given by chart entry times of songs and we assume that musical similarity correlates with chart entry time proximity. Whereas song year prediction has received little attention in the literature, the song year is important in determining musical preference \cite{barrett2010music}. Thus, song year prediction might be applied in music recommendation \cite{bertin2011million}. Song year prediction might furthermore be incorporated in genre classification tasks, since musical genres are associated with particular years. 

Section~\ref{sec:background} provides an overview of methods and descriptors for computing low-specificity similarity. In Section~\ref{sec:approach}, we describe our approach. In Section~\ref{sec:evaluation}, we detail our experimental method and results; we provide separate accounts for similarity rating prediction and song year prediction in Sections~\ref{sec:evaluationsimilarityratingprediction} and \ref{sec:evaluationsongyearprediction}, respectively. Finally, in Section~\ref{sec:conclusions} we provide conclusions. 

\section{Background}
\label{sec:background}
For a detailed review of recent literature on methods for determining musical similarity, from the perspective of classification, we refer to the work of Fu et al.~\cite{fu2011survey}. To determine musical similarity, one possible approach involves computing pairwise distances between tracks. The obtained distances may then be used for classification. A second approach consists in applying track-wise descriptors directly for classification.

Based on the second approach, Tzanetakis and Cook~\cite{tzanetakis2002musical} compute first and second-order moments on spectral features including MFCCs, to perform genre classification using the $k$-nearest neighbours (KNN) algorithm and Gaussian mixture models (GMMs) estimated on each target class. Li and Ogihara \cite{li2006toward} propose to classify Daubechies wavelet histograms using GMMs and KNN for genre and mood classification. Using spectral features, West et al.~\cite{west2006incorporating} propose methods for learning similarity functions based on constructing decision trees for genre classification. Slaney et al.~\cite{slaney2008learning} propose feature transformations based on supervised learning and using onset and loudness features, for the purpose of album and artist classification.

Based on the approach of determining distances between descriptors, Logan and Salomon~\cite{logan2001music} propose to estimate GMMs on individual tracks. Pairwise track distances are then computed using a combination of Kullback-Leibler divergence (KLD) and earth mover's distance, where the KLD is used to compare pairs of track centroids. The approach based on KLD assumes that each centroid follows a Gaussian distribution; thus the KLD may be computed in closed form as
\begin{equation}
\label{eqn:kldivergence}
\begin{split}
\mathrm{KLD} = & \frac{1}{2} \left( \vphantom{\frac{|\boldsymbol{\Sigma}_2|}{|\boldsymbol{\Sigma}_1|}} \mathrm{tr} \left(\boldsymbol{\Sigma}_1^{-1} \boldsymbol{\Sigma}_2 \right)  + \left(\greekvec{\mu}_1 - \greekvec{\mu}_2 \right)^T \boldsymbol{\Sigma}_1^{-1} \left(\greekvec{\mu}_1 - \greekvec{\mu}_2 \right) \right. \\ & \left.  - h - \log \frac{|\boldsymbol{\Sigma}_2|}{|\boldsymbol{\Sigma}_1|} \right)
\end{split}
\end{equation}
where $\boldsymbol{\Sigma}_1, \boldsymbol{\Sigma}_2$ and $\greekvec{\mu}_1, \greekvec{\mu}_2$ respectively denote the mean and covariance of two multivariate Gaussian distributions with dimensionality $h$. Aucouturier and Pachet \cite{aucouturier2002music} in contrast compute cross-likelihoods between GMMs using Monte Carlo approximations for the purpose of genre classification, whereas Berenzweig et al.~\cite{berenzweig2004large} consider the asymptotic likelihood approximation of the KLD and centroid distances for the task of similarity rating prediction. Mandel and Ellis~\cite{mandel2005song} instead represent tracks as single Gaussians and use (\ref{eqn:kldivergence}) as a distance measure between track pairs. The obtained distances are then applied to artist identification, using support vector machines (SVMs) for classification. An alternative approach to computing the KLD is based on computing histograms of quantised features, as proposed by Vignoli and Pauws~\cite{vignoli2005music} for playlist recommendation; Levy and Sandler \cite{levy2006lightweight} compare approaches in the context of genre classification.

The previously described techniques are commonly referred to \textit{bag-of-features} approaches, since they discard information on temporal structure. Yet, the relative convenience of bag-of-features approaches stands in contrast to the importance of temporal structure in perception of musical timbre, as observed by McAdams et al.~\cite{mcadams1995perceptual}. Aucouturier and Pachet \cite{aucouturier2007bag} argue that the bag-of-features approach is insufficient to model polyphonic music for determining similarity. Sequential representations based on mid-level features are widely applied for the purpose of version identification \cite{serra2011identification}. For low-specificity classification, one possible approach to mitigating the shortcoming of the \textit{bag-of-features} approach involves the intermediate step of aggregating features locally, before summarising anew using obtained summary statistics. Tzanetakis and Cook~\cite{tzanetakis2002musical} propose to estimate the local mean and variance of features contained in a 1s window. For the task of predicting musical similarity, Seyerlehner et al.~\cite{seyerlehner2010fusing} apply a single, global summarisation step to overlapping windows, computing variance and percentiles. For the purpose of local aggregation, alternative pooling functions are considered by M{\"o}rchen et al.~\cite{morchen2006modeling}, Hamel et al.~\cite{hamel2011temporal}, W{\"u}lfing and Riedmiller~\cite{wulfing2012unsupervised}.

An alternative approach is based on retaining the temporal order of features at each window position. Spectral analysis may be applied to the original features, resulting in a new feature sequence. Pampalk \cite{pampalk2006computational} proposes fluctuation patterns describing loudness modulation across frequency bands, whereas Lee et al.~\cite{lee2009automatic} propose statistics based on modulation spectral analysis. M{\"o}rchen et al.~\cite{morchen2006modeling} consider a variety of statistics based on spectral analysis and autocorrelation. Meng et al.~\cite{meng2007temporal}, Coviello et al.~\cite{coviello2012multivariate} apply multivariate autoregressive modelling to windowed features, for the tasks of genre and tag classification.

To account for temporal structure, statistical modelling may be applied to quantised features. For genre classification, Li and Sleep~\cite{li2005genre} propose an SVM kernel in which pairwise distances are obtained by comparing dictionaries generated using the Lempel-Ziv compression algorithm \cite{ziv1978compression}. Reed and Lee \cite{reed2009importance} apply latent semantic analysis to unigram and bigram counts for classification using SVMs, whereas Langlois and Marques \cite{langlois2009music} propose to estimate language models for computing sequence cross-likelihoods for genre and artist classification. Ren and Jang~\cite{ren2012discovering} propose an algorithm for computing histograms of feature codeword sequences for genre classification.

Recent approaches attempt to model temporal structure using representations constructed at multiple time scales. Based on a bag-of-features approach, Foucard et al.~\cite{foucard2011multi} propose an ensemble of classifiers, where each classifier is trained on features at a given time scale. Features at successive resolutions are aggregated using averaging. Applied to tag and instrument classification, results indicate that a multiscale approach benefits  performance. Dieleman and Schrauwen~\cite{dielemanmultiscale} apply feature learning based on spherical $K$-means clustering to tag classification. Evaluated  aggregation techniques are based on varying the spectrogram window size, in addition to Gaussian and Laplacian pyramid smoothing techniques. Although not applied to classification, Mauch and Levy~\cite{mauch2011structural} propose a similar smoothing approach for characterising structural change at multiple time scales. Finally, convolutional neural networks have been proposed for modelling temporal structure: Dieleman et al.~\cite{dieleman2011audio} propose deep learning architectures for genre, artist and key classification tasks. Hamel et al.~\cite{hamel2011temporal} propose a deep learning architecture incorporating multiple feature aggregation functions for tag classification.

The approach proposed in this work resembles methods applying statistical models to quantised feature sequences \cite{li2005genre,reed2009importance,langlois2009music,ren2012discovering}. In contrast, we propose to compute summary statistics directly from estimated sequential models. Since the obtained statistics may be compared using a metric, our approach has the potential to be combined with indexing and hashing schemes for computationally efficient retrieval \cite{slaney2008locality,rhodes2010investigating,schluter2013}, while retaining information on temporal structure. Our method of computing multiple representations using downsampling resembles the approach proposed by Dieleman and Schrauwen \cite{dielemanmultiscale}.

Note that our approach differs from Cilibrasi et al.~\cite{cilibrasi2004algorithmic}, who propose pairwise sequence compressibility to quantify similarity. We did not pursue this approach for low-specificity tasks, based on results for the pairwise prediction approach reported in Section~\ref{sec:similarityratingpredictionresults}. Note that we may take compression rates as estimates of sequential Shannon entropy rates, inviting further comparison or combination with related measures of sequential complexity \cite{dubnov2008unified,abdallah2009information,james2011anatomy}. Such measures have to date not been evaluated quantitatively in music content analysis, inviting further investigation beyond the scope of this work.

\section{Approach}
\label{sec:approach}
Assume that we are given the audio feature vector sequence $\mathbf{V} = \left(\vec{v}_1, \ldots, \vec{v}_T \right)$. Similar to the descriptor proposed in \cite{streich2005automatic}, as a means of quantifying the sequential complexity of $\mathbf{V}$, we compute the compression rate $R_\lambda(\mathbf{V})$,
\begin{equation}
\label{eqn:compressionrate}
R_\lambda(\mathbf{V}) = \frac{C(\mathbf{V}, \lambda)}{T}
\end{equation}
where $C(\mathbf{V}, \lambda)$ denotes the number of bits required to represent $\mathbf{V}$, given a quantisation scheme with $\lambda$ levels and using a specified sequential compression scheme. To obtain a length-invariant measure of sequential complexity, we normalise with respect to the sequence length $T$.

Given the $i$th track in our collection, we compute compression rates for feature sequences extracted from musical audio. We refer to the set of compression rates as \textit{feature complexity descriptors} (FCDs). For features based on constant frame rate, we compute FCDs using the original feature sequence, in addition to FCDs computed on downsampled versions of the original sequence; we consider downsampling factors ${1, 2, 4, 8}$. We distinguish among temporal resolutions using the labels FCD1, FCD2, FCD4, FCD8, respectively. For features based on variable frame rate, we compute FCDs with no further downsampling applied. 

\begin{figure}[h]
  \begin{minipage}{.31\linewidth}
	\centering
    \centerline{\includegraphics{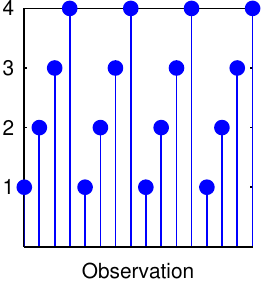}}
	\centerline{(a) Low $R_\lambda$}
  \end{minipage}
  \begin{minipage}{.31\linewidth}
	\centering
    \centerline{\includegraphics{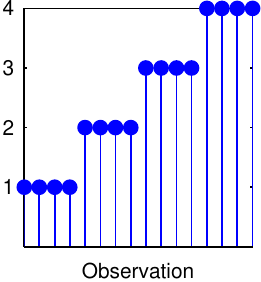}}
	\centerline{(b) Low $R_\lambda$}
  \end{minipage}
  \begin{minipage}{.31\linewidth}
	\centering
    \centerline{\includegraphics{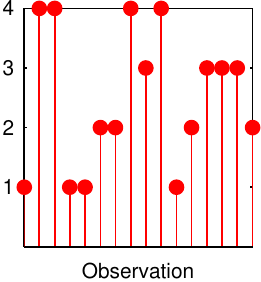}}
	\centerline{(c) High $R_\lambda$}
  \end{minipage}
	\caption{Hypothetical sequences with low and high $R_\lambda$, assuming $\lambda = 4$.}
    \label{fig:examplesequences}
\end{figure}

Thus proposed, consider FCDs computed on a hypothetical scalar-valued feature sequence exhibiting a high amount of temporal structure, either due to periodicity or locally constant regions (Fig.~\ref{fig:examplesequences} (a), (b)). For such sequences, we obtain low values for $R_\lambda$, since the quantised feature sequence may be encoded efficiently. Conversely, if we discard temporal structure by randomly shuffling the original feature sequence (Fig.~\ref{fig:examplesequences} (c)), we obtain high values for $R_\lambda$, since the quantised feature sequence no longer admits an efficient encoding. In contrast to FCDs, feature moments such as mean and variance are invariant to any such re-ordering of features. We observe that feature moments have been widely applied for low-specificity content analysis tasks. Considering that FCDs have similar dimensionality to feature moments and assuming that temporal order of features is informative for our considered tasks, we therefore expect that FCDs may be used to improve prediction accuracy with respect to using feature moments alone, for our considered tasks.

\subsection{Similarity rating prediction}
\label{sec:predictingsimilarityratings}
For the task of similarity rating prediction, assume that we have a distance metric which we use to compare descriptor vectors computed on pairs of tracks. We hypothesise that the pairwise distance between descriptors correlates with the similarity rating associated with track pairs.
To predict similarity ratings we take as our feature space pairwise distances between descriptor vectors and apply multinomial regression. 
We use $\vec{r}_{i,n}$ to denote the $n$th descriptor vector computed for the $i$th track in our collection, with $1 \leq n \leq N$ and given a set of $N$ available descriptor vectors. We compute separate descriptor vectors across audio features and across FCD resolutions, with each vector component in $\vec{r}_{i,n}$ corresponding to a quantisation granularity $\lambda$. We denote with $\vec{d}_{\langle i, j \rangle}$ the distances between $\vec{r}_{i,n}$, $\vec{r}_{j,n}$ obtained across all $N$ descriptor vectors, using our assumed distance measure. Given the pair of tracks $\langle i, j \rangle$ whose similarity rating we seek to predict, we estimate the probability of similarity score $k \in [1\,..\,K]$ as
\begin{equation}
\label{eqn:multinomialregressionmodel}
  P\left(S=k | \vec{d}_{\langle i, j \rangle}\right) = \frac{\exp \left( \greekvec{\beta}_{k}^T \, \vec{d}_{\langle i, j \rangle} + \gamma_k \right)}{\sum_{m=1}^K \exp\left( \greekvec{\beta}_{m}^T \, \vec{d}_{\langle i, j \rangle} + \gamma_m   \right)}
\end{equation}
where $\greekvec{\beta}_k$, $\gamma_k$ are the model parameters associated with outcome $k$, given a total of $K$ similarity scores. We predict similarity ratings by determining the value of $k$ which maximises $P\left(S=k | \vec{d}_{\langle i, j \rangle}\right)$. We describe our model estimation method in Section~\ref{sec:modelestimation}. 

\subsection{Song year prediction}
\label{sec:predictivemodels}
For the task of song year prediction, we hypothesise that descriptor values correlate with the chart entry date of tracks. Following \cite{bertin2011million} we apply a linear regression model. Given the $i$th track in our collection, we predict the associated chart entry date ${y}_i$ using a linear combination of components in descriptor vectors $\vec{r}_{i,n}$,
\begin{equation}
\label{eqn:regressionmodel}
\hat{y}_i = \sum_{n=1}^{N}  \boldsymbol{\theta}_n^T \vec{r}_{i,n} + \alpha
\end{equation}
where $\boldsymbol{\theta}_n$ denotes regression  coefficients for the $n$th descriptor vector as specified for similarity rating prediction, and where $\alpha$ denotes the model intercept. We describe our model estimation method for song year prediction in Section~\ref{sec:modelestimationyearprediction}. We motivate use of both multinomial and linear regression techniques as a straightforward means of evaluating the utility of FCDs for determining similarity based on a metric space. We perform our evaluation by considering predictive accuracy, in addition to interpreting estimated coefficients as feature utilities.  

\section{Evaluation} 
\label{sec:evaluation}
For our evaluations, we use a collection of 15\,473 entries from the American \textit{Billboard Hot 100} singles popularity chart\footnote{\url{http://www.billboard.com}}. Each entry in the dataset is represented by a track excerpt of approximately 30s of audio, and is annotated with a chart entry date. Chart entry dates span the years 1957--2010 ($\textit{M} = 1982.9$y, $\textit{SD} = 15.4$y). 

For each track excerpt in the dataset, we extract a set of 25 audio features, using MIRToolbox \cite{lartillot2007matlab} version 1.3.2 and using the framewise chromagram representation proposed by Ellis and Poliner \cite{ellis2007identifyingcover}. With the exception of rhythmic features, which are computed using predicted onsets, features are based on a constant frame rate of $40$Hz. Table~\ref{tab:featuresummary} summarises the set of evaluated audio features.

\begin{table}[th]
\centering
\opt{journal}{\begin{tabular}{p{3cm} p{5.0cm}}}
\opt{preprint}{\begin{tabular}{p{0.4\linewidth} p{0.6\linewidth}}}
\hline
\hline
Feature name & Description \\
\hline
Chroma (Ellis and Poliner)  & 12-component chromagram based on using phase-derivatives to identify tonal components in spectrum \cite{ellis2007identifyingcover}.\\
dynamics.rms & Root mean square of amplitude.\\
rhythm.tempo & Tempo estimate based on selecting peaks from autocorrelated onsets.\\
rhythm.attack.time & Duration of onset attack phase.\\
rhythm.attack.slope & Slope of onset attack phase.\\
spectral.centroid & First moment of magnitude spectrum.\\
spectral.brightness & Proportion of spectral energy above 1500Hz.\\
spectral.spread & Second moment of magnitude spectrum.\\
spectral.skewness & Skewness coefficient of magnitude spectrum.\\
spectral.kurtosis & Excess kurtosis of magnitude spectrum.\\
spectral.rolloff95 & 95th percentile of energy contained in magnitude spectrum.\\
spectral.rolloff85 & 85th percentile of energy contained in magnitude spectrum.\\
spectral.spectentropy & Shannon entropy of magnitude spectrum.\\
spectral.flatness & Wiener entropy of magnitude spectrum.\\
spectral.roughness & Average roughness \cite{plomp1965tonal} between peak pairs in magnitude spectrum.\\
spectral.irregularity & Squared amplitude difference between successive partials \cite{jensen1999timbre}.\\
spectral.mfcc & 12-component MFCCs \cite{slaney1998auditory} (excluding energy coefficient).\\
spectral.dmfcc & First-order differentiated MFCCs.\\
spectral.ddmfcc & Second-order differentiated MFCCs.\\
timbre.zerocross & Zero crossing rate.\\
timbre.spectralflux & Half-wave rectified L1 distance between magnitude spectrum at successive frames \cite{masri1996computer}.\\
tonal.chromagram.centroid & Centroid of 12-component chromagram.\\
tonal.keyclarity & Peak correlation of chromagram with key profiles \cite{gomez2006tonal}.\\
tonal.mode & Predicted mode after correlating chromagram with key profiles.\\
tonal.hcdf & Flux of 6-dimensional tonal centroid \cite{harte2006detecting}.
\end{tabular}
\caption{Summary of evaluated audio features.}
\label{tab:featuresummary}
\end{table}

\begin{table*}[h]
\centering
\begin{tabular}{c l l c l l}
\hline
\hline
\multicolumn{3}{c}{\scriptsize Lowest-ranking scores} & \multicolumn{3}{c}{\scriptsize Highest-ranking scores}\\
Score & Artist name & Medoid track name & Score & Artist name & Medoid track name\\
\hline
1.223 & Johnny Mathis & Starbright & 1.286 & Jan  \&  Dean & The Anaheim \ldots Association\\
1.234 & Barbra Streisand & Didn't We &   1.286 & Bryan Adams & This Time\\
1.240 & The Platters & Trees &1.287 & Eric Clapton & After Midnight\\
1.245 & Bobby Vinton & Rain Rain Go Away   & 1.287 & Creedence Clearwater Revival & Who'll Stop The Rain\\
1.247 & Connie Francis & (He's My) Dreamboat  &  1.287 & The Rolling Stones & Tell Me (You're Coming Back)\\
1.251 & Andy Williams & Sweet Memories & 1.288 & Johnny Cash & It's Just About Time\\
1.252 & Jim Reeves & I Guess I'm Crazy & 1.288 & Chubby Checker & Whole Lotta Shakin' Goin' On\\
1.256 & John Denver & Sweet Surrender  & 1.288 & The Kinks & Tired Of Waiting For You\\
1.256 & Barry Manilow & I Write The Songs &  1.288 & Eddie Money & Maybe I'm A Fool\\
1.261 & Johnny Tillotson & I Rise, I Fall &  1.288 & Aerosmith & Hole In My Soul\\
1.261 & Dionne Warwick & If We Only Have Love &  1.288 & Van Halen & When It's Love\\
1.261 & Helen Reddy & Delta Dawn   & 1.289 & The Doobie Brothers & What A Fool Believes\\
1.262 & Etta James & Seven Day Fool &1.289 & Marvin Gaye & Pretty Little Baby\\
1.263 & Carpenters & Touch Me When We're Dancing  &  1.289 & Madonna & Secret\\
1.263 & Frank Sinatra & Talk To Me & 1.290 & Paul Revere  \&  The Raiders & Country Wine\\
1.264 & Engelbert Humperdinck & In Time &1.291 & James Brown & Signed, Sealed, And Delivered\\
1.264 & Brenda Lee & Too Many Rivers  &  1.291 & Janet Jackson & Black Cat\\
1.264 & Nat King Cole & Nothing In The World &   1.291 & The Isley Brothers & Harvest For The World\\
1.266 & Gene Pitney & Town Without Pity & 1.293 & Freddy Cannon & Muskrat Ramble\\
1.267 & Tom Jones & With These Hands    & 1.297 & Eminem & Cleanin' Out My Closet\\

\end{tabular}
\caption{Artists ranked according to median track-wise FCD score. For each artist, FCDs averaged across quantisation levels $\lambda$ and across temporal resolutions, using MFCCs as audio feature. Table reports lowest-ranking and highest-ranking scores.}
\label{tab:rankedcomplexities}
\end{table*}

In addition to FCDs, for each track excerpt we compute the mean and standard deviation, based on frame-level representation with no downsampling applied. We refer to the latter non-sequential descriptors as feature moment descriptors (FMDs).
We compute FCDs as described in Section~\ref{sec:approach}, where for the case of the vector-valued features chroma, MFCCs and delta-MFCCs we apply principal component analysis (PCA) in track-wise fashion as a preliminary decorrelation step. We then quantise and compress each resulting component separately, before averaging obtained compression lengths across components. We apply PCA, since we seek to quantify temporal structure in feature vector sequences while disregarding any correlation among feature vector components. We quantise features by applying equal-frequency binning with $\lambda \in \{3, 4, 5\}$ levels; we perform relatively coarse quantisation to ensure that each symbol occurs frequently, regardless of downsampling factor.

We choose equal-frequency binning to ensure that obtained strings have a consistent stationary distribution; the obtained compression rates therefore are a function of temporal structure alone. The value $\log \lambda$ may be interpreted as the theoretical compression rate for a temporally uncorrelated sequence. We compress symbol sequences using the \textit{prediction by partial match} (PPM) algorithm\footnote{\url{http://www.cs.technion.ac.il/~ronbeg/vmm/index.html}}, described in \cite{begleiter2004prediction}. We consider PPM a general-purpose string compression algorithm which may be substituted with an alternative compressor; in initial experiments we obtained similar results using Lempel-Ziv compression \cite{ziv1978compression}. Nevertheless, we note that PPM compresses efficiently compared to alternative compression schemes \cite{begleiter2004prediction}. We set the PPM model order to 5 symbols, based on the observation that for uncorrelated sequences, distinct substrings of length 5 are unlikely to occur frequently.

With a view to characterising the feature space represented by FCDs, we perform a track-wise exploratory analysis of computed FCDs. For each track excerpt in our collection, we compute FCDs based on MFCC features alone. We obtain a scalar-valued score for each excerpt by averaging FCDs across quantisation levels $\lambda$ and across temporal resolutions. Next, across artists in our collection we compute the median of obtained FCD scores. To facilitate interpretation, we consider only artists with a minimum number of 20 chart entries; thus out of 5\,455 artists in our collection we consider 129 artists. We then rank artists according to median FCD scores. Shown in Table~\ref{tab:rankedcomplexities}, we report the 20 lowest-ranking and highest-ranking artists. Additionally, across artists we report as medoid tracks those tracks whose FCD score minimises the error with respect to the median. 

Comparing track groups, the lowest-ranking artists are predominantly vocalists with repertoire of jazz ballads and slow-moving pieces (e.g. Johnny Mathis, Barbara Streisand). In contrast, the artists with highest complexity values stand for music with strong percussive and aggressive components, from up-tempo surf-rock (Jan \& Dean), through 1980s Power Rock (Van Halen) and Hip Hop (Eminem). Informal listening to medoid tracks supports this observation, with the exception of the medoid track by artist Etta James. We view this observation in support of our expectation that FCDs may be useful for low-specificity similarity and subsequently demonstrate validity of our expectation for the similarity tasks considered in this work. Note however that we make no claim that FCDs capture any notion of musical complexity as proposed in \cite{pressing1999cognitive}. While beyond the scope of this paper, track-wise analysis of FCDs merits further investigation.

\subsection{Similarity rating prediction}
\label{sec:evaluationsimilarityratingprediction}
We evaluate similarity rating prediction using annotations collected for a subset of the chart music dataset. Prior to our investigations, we obtained a total of 7\,784 pairwise similarity ratings from 456 subjects participating in a web-based listening test\footnote{\url{http://webprojects.eecs.qmul.ac.uk/matthiasm/audioquality-pre/check.php}}. Subjects were asked to quantify pairwise musical similarity between successive pairs of track excerpts using a five-point ordinal scale, with score `1' corresponding to `not similar' and score `5' corresponding to `very similar'. We assume that subjects have an internal similarity scale which they use to perform ratings. Therefore, we omit any training step from the rating process. Note that while we prescribe that pairwise similarity ratings are made using a five-point scale, we do not assume that similarities are judged using an absolute scale across listeners. Given three track pairs for which we have respective ratings (4,~5), (5,~5), (1,~2), we view the ratings as quantifying relative agreement, compared to (4,~1), (5,~1), (1,~4).

For human similarity judgements, two issues prompt consideration: In addition to music being inherently subjective \cite{wiggins2010non}, human similarity judgements are context-dependent \cite{goodman1972seven, tversky1977features}.
We motivate our assumption of an internal similarity scale on the basis that Western popular music is widely disseminated and that listeners might form similarity judgements using a common factor. We verify our assumptions by quantifying similarity rating agreement. 

When presenting track pairs to listeners, we select the first song in each pair using uniform sampling. For the second song in each pair, we again apply uniform sampling, however we bias towards proximate chart entry times by restricting the permissible chart entry deviation to $\leq 1$y with probability $0.9$. We bias as a means of controlling for historical changes in audio production, which might affect similarity ratings \cite{sturm2012two}.
We obtain a median of 6 ratings per subject, with each rating corresponding to a unique track pair. Table~\ref{tab:ratinglevelhistogram} displays obtained score counts.

As shown in Table~\ref{tab:ratinglevelhistogram}, the majority of ratings are associated with scores less than `3', corresponding to relative dissimilarity on the five-point scale. We contend that for music content analysis based on an ensemble of systems as proposed in \cite{bogdanov2011unifying}, the entire target set of predicted musical similarity might be used when forming recommendations. In contrast, for track recommendation relying on predicted similarity alone, when forming recommendations, it is typically of interest to consider tracks deemed similar to a query, while disregarding tracks deemed dissimilar \cite{downie2010music}. Pertaining to the first use case, we perform evaluations using the five-point scale ratings, as defined previously. Pertaining to the second use case, we merge similarity ratings with scores `1' and `2', thus discarding any distinction between similarity ratings with low scores. We then perform our evaluations using the resulting four-point scale ratings.

\begin{table}[h]
\centering
\begin{tabular}{c| c c c c c}
  & \multicolumn{5}{c}{\scriptsize{Similarity score}} \\
  & \textbf{1} & \textbf{2} & \textbf{3} & \textbf{4} & \textbf{5} \\
\hline
Count & 2060 & 2115 & 1742 & 1391 & 476 \\
\end{tabular}
\caption{Similarity score counts obtained from web-based listening test.}
\label{tab:ratinglevelhistogram}
\end{table}

To assess the consistency of similarity ratings, we collected an additional set of similarity ratings under controlled experimental conditions, involving 12 subjects aged 21y--42y. Subjects were assessed using the Ollen musical sophistication index (OMSI) \cite{ollen2006criterion}. We obtain a median OMSI score score of 241, with an associated median of 0.75 years of formal musical training. To avoid subject fatigue, we imposed no minimum number of ratings per subject, and collected ratings during two 30-minute sessions. We selected stimuli by sampling uniformly from the set of track pairs for which we have prior ratings. Across subjects, we obtain a median of 42 ratings ($\textit{M} = 45.4$, $\textit{SD} = 29.3$). We aggregate controlled-condition ratings across subjects and thus obtain a total of 509 controlled-condition similarity ratings, corresponding to $6.5\%$ coverage of web-based similarity ratings. Table~\ref{tab:ratingagreementconfusionmatrix} displays a confusion matrix of web-sourced versus controlled-condition similarity ratings.

\begin{table}[th]
\centering
\begin{tabular}{c c | c c c c c}
 & & \multicolumn{5}{c}{\scriptsize Controlled-condition} \\
 & & \textbf{1} & \textbf{2} & \textbf{3} & \textbf{4} & \textbf{5} \\
\hline
\multirow{5}{*}{\rotatebox{90}{\scriptsize Web-sourced~}}
 & \textbf{1} & 64 & 34 & 17 & 10 & 0 \\
 & \textbf{2} & 55 & 44 & 18 & 14 & 4 \\
 & \textbf{3} & 26 & 41 & 26 & 25 & 5 \\
 & \textbf{4} & 16 & 30 & 16 & 24 & 7 \\
 & \textbf{5} & 6 & 9 & 5 & 8 & 5 \\
\end{tabular}
\caption{Confusion matrix of web-sourced versus controlled-condition similarity ratings.}
\label{tab:ratingagreementconfusionmatrix}
\end{table}

We quantify the agreement between controlled-condition and web-sourced similarity ratings. We report results for both five-point and four-point rating scales; for each agreement statistic we report results for the four-point rating scale in brackets. We first quantify agreement using Kendall's correlation coefficient $\tau_b$, as defined in (\ref{eqn:kendalltau}).
We obtain a correlation of $0.274$ ($0.250$), with $p < 0.001$ based on a permutation test for the hypothesis of no correlation. We then compute a confidence interval for the obtained sample correlation by applying bootstrap sampling \cite{efron1982jackknife}. At the 95\% level, we obtain correlations in the range $[0.205, 0.337]$ ($[0.173, 0.325]$). Subsequently, we consider the correlation $0.337$ ($0.325$) an upper bound on attainable accuracy using our proposed method of similarity rating prediction. As a second measure of rating agreement, we compute Spearman's correlation coefficient $\rho_s$, where we obtain $0.329$ ($0.278$) for ratings aggregated across subjects. Analogously by applying bootstrap sampling, at the 95\% level we obtain correlations in the range $[0.247, 0.404]$ ($[0.193, 0.361]$). We consider the correlation $0.404$ ($0.361$) an upper bound on attainable accuracy based on $\rho_s$. Finally, using Table~\ref{tab:ratingagreementconfusionmatrix} and interpreting the controlled-condition rating process as a multinomial classification task, we obtain a balanced classification accuracy (BA) of $0.292$ $(0.345)$; the corresponding 95\% confidence interval is $[0.254, 0.336]$ ($[0.304, 0.393]$).

\subsubsection{Distance measures}
\label{sec:distancemeasures}
We predict similarity ratings by applying multinomial regression to pairwise Euclidean distances between descriptor vectors, using the approach described in Section~\ref{sec:predictingsimilarityratings}. As an additional baseline distance measure, using (\ref{eqn:kldivergence}) and assuming Gaussianity and diagonal covariance, we compute the KLD on pairs of FMDs. We logarithmically transform distances obtained using the KLD, which we observed improved prediction accuracy.

As a baseline distance accounting for temporal structure, we compute the cross-prediction error between audio feature sequences, with each feature sequence represented at the original frame level. Following \cite{serra2012predictability}, we apply state space embedding \cite{takens1981detecting} separately to pairs of feature sequences. Given feature vectors $(\vec{v}_1, \ldots, \vec{v}_T)$ each with dimensionality $h$, state space embedding produces higher-dimensional feature vectors with dimensionality $d h$ by stacking $d$ consecutive vectors $\vec{v}_{t-d}, \ldots \vec{v}_{t-1}$ at each time step $t$. We perform cross-predictions by determining sequential successors of nearest neighbours in the embedded space, using the approach given in \cite{foster2013identification}. As a distance measure between predicted and observed feature sequences, we compute the normalised mean square error \cite{serra2012predictability}. We consider parameter $d \in \{8, 12, 16, 20 \}$ and report results for $d = 12$, which yields highest average correlation between computed pairwise distances and similarity annotations. We apply square-root transformation to pairwise distances, which we observed improved similarity rating prediction accuracy.

\subsubsection{Performance statistics}
\label{sec:performancestatistics}
To quantify the accuracy of similarity rating prediction, as discussed in \cite{cardoso2011measuring} we compute Kendall's $\tau_b$ and Spearman's $\rho_s$, both which are ordinal measures of association between predicted and annotated similarity ratings. We define Kendall's $\tau_b$ as follows. Assume that we have sequences $\mathcal{Q} = (q_1, \ldots, q_M)$, $\mathcal{O} = (o_1, \ldots, o_M)$. The pair $d_{i,j} = ((q_i,o_i),(q_j, o_j))$ is termed \textit{concordant}, if $q_i > q_j$ and $ o_i > o_j$, or if $q_i < q_j$ and $o_i < o_j$. Analogously, $d_{i,j}$ is termed \textit{discordant}, if $q_i < q_j$ and $o_i > o_j$, or if $q_i > q_j$ and $o_i < o_j$.
Kendall's $\tau_b$ is defined as
\begin{equation}
  \label{eqn:kendalltau}
  \tau_b = \frac{M_c - M_d}{\sqrt{(M_p - M_q) (M_p - M_o)}}
\end{equation}
where $M_c$, $M_d$ respectively denote the number of concordant and discordant pairs and where $M_p = \frac{1}{2} M (M-1)$ denotes the total number of pairs. Terms $M_q$, $M_o$ respectively denote the number of pairs with tied $(q_i, q_j)$ and with tied $(o_i, o_j)$. In the denominator, the normalisation is with respect to the geometric mean of adjusted pair counts $(M_p - M_q)$, $(M_p - M_o)$.
Yielding values in the range $[-1,1]$, $\tau_b$ may be interpreted as an estimate of the difference in probability of sampling a concordant pair versus sampling a discordant pair in $(\mathcal{Q}, \mathcal{O})$, while accounting for ties.

As a second measure of prediction accuracy, we compute Spearman's $\rho_s$, corresponding to the product-moment correlation coefficient between separately ranked $\mathcal{Q}$, $\mathcal{O}$. We assign unique ranks to tied values, before computing average ranks across tied values. Note that in contrast to $\tau_b$, the value of $\rho_s$ is a function of assigned ranks. Thus, in the presence of ties $\tau_b$ may be viewed as a more appropriate means of comparing ordinal sequences \cite{pinto2008unimodal}. Nevertheless, we compute $\rho_s$, since its square yields a direct interpretation as proportion of explained variance between assigned ranks. 

As a third performance measure, we view our prediction task as multinomial classification and compute BA. Note that in contrast to $\tau_b$, $\rho_s$, BA disregards the ordering of rating scores. Based on the notion of rating agreement given in Section~\ref{sec:evaluationsimilarityratingprediction}, we thus consider BA a subsidiary measure of performance compared to $\tau_b$, $\rho_s$.

\subsubsection{Model estimation}
\label{sec:modelestimation}
We evaluate similarity rating prediction by applying hold-out validation to web-sourced annotations. We use 60\% of annotations for training, with the remainder of annotations used for testing.

We apply multinomial regression separately to sets of distances between descriptor vectors, as specified in Table~\ref{tab:similarityprediction}. We standardise distances  by subtracting the mean and dividing by the variance of the training data. Note that we compute FCD vectors separately across temporal resolutions and across audio features. Based on a set of 25 audio features, given a pair of tracks we thus obtain a total of 100 distances between compression-based descriptor vectors. Furthermore, note that when combining sets of descriptors we aggregate among obtained distances. Thus given a pair of tracks, when combining sets 1, 3, 4 as specified in Table~\ref{tab:similarityprediction}, we obtain 150 distances. As given in (\ref{eqn:multinomialregressionmodel}), we weight distances individually.

In our training step, we estimate multinomial regression parameters using elastic net regularisation (ENR) \cite{zou2005regularization} based on coordinate descent\footnote{\url{http://www.stanford.edu/~hastie/glmnet_matlab/}} \cite{friedman2010regularization}. We denote with  $\greekvec{\beta} = (\greekvec{\beta}_1^T, \ldots, \greekvec{\beta}_K^T )^T$, $\greekvec{\gamma} = (\gamma_1, \ldots, \gamma_K)^T$  regression coefficients and model intercepts as given in~(\ref{eqn:multinomialregressionmodel}). Using ENR, we solve
\begin{equation}
\label{eqn:enrmultinomialregression}
  \min_{\greekvec{\beta}, \greekvec{\gamma}} \left\{ \eta \left( \nu \|\greekvec{\beta} \|_1 +  \left(1 - \nu \right) \frac{1}{2} \| \greekvec{\beta} \|_2^2 \right) - \ell(\greekvec{\beta}, \greekvec{\gamma}) \right\}
\end{equation}
where $\ell(\greekvec{\beta}, \greekvec{\gamma})$ denotes model log-likelihood. Furthermore,  $\eta$ and $\nu$ respectively are \textit{shrinkage} and \textit{elastic net penalty}  parameters, with $\eta > 0$ and $0 \leq \nu \leq 1$. Thus, $\nu$ determines the relative contribution of regularisation due to L1 and L2 norms, whereas $\eta$ scales the regularisation penalty. For each performance statistic given in Section~\ref{sec:performancestatistics} and for each rating scale as given in \ref{sec:evaluationsimilarityratingprediction}, we apply hold-out validation to training data and optimise $\eta$ by determining maximal prediction accuracy. We consider $\nu$ a hyper-parameter which we assign constant value; we optimise Kendall's $\tau_b$ with respect to the five-point rating scale and using a model incorporating FCDs and FMDs, where we again apply hold-out validation to training data.

\begin{table*}[th]
\centering
\opt{journal}{\begin{tabular}{c l l l l}}
\opt{preprint}{\begin{tabular}{p{0.025\linewidth} p{0.26\linewidth} p{0.22\linewidth} p{0.20\linewidth} p{0.14\linewidth}}}
\hline
\hline
Set & Track representation & Descriptor vector components & Distance measure & Prediction coeffs.\\
\hline
1. & FCDs               &  $\lambda \in \{3, 4, 5 \}$ & Euclidean & $4 \times 25 $ \\
2. & Frame sequence     & N/A             & Cross-prediction error & $25$ \\
3. & FMDs               & Mean, Std       & Euclidean & $25$ \\
4. & FMDs               & Mean, Var       & KLD                & $25$ \\
5. & Combine 3, 4       &                 &                                 & $50$ \\
6. & Combine 1, 3, 4       &                 &                                 & $150$
\end{tabular}
\caption{Summary of descriptor combinations evaluated for similarity rating prediction. Third column denotes components included in descriptor vectors. Fifth column lists number of coefficients in multinomial regression model (excluding intercepts).}
\label{tab:similarityprediction}
\end{table*}

\begin{table*}[th]
\centering
\begin{tabular}{c l l l l}
\hline
\hline
Set & Track representation & Descriptor vector components & Prediction coeffs.\\
\hline
1. & FMDs               & Mean, Std                       & $21\times 2 + 4 \times 24$\\
2. & FCDs & $\lambda \in \{3, 4, 5 \}$ & $25 \times 4 \times 3$ \\
3. & Combine 1, 2       &                                                   & 
\end{tabular}
\caption{Summary of descriptor combinations evaluated for song year prediction. Fourth column lists number of coefficients in linear regression model (excluding intercept).}
\label{tab:yearprediction}
\end{table*}

\subsubsection{Results}
\label{sec:similarityratingpredictionresults}

\begin{figure}[h]
    \centering
    \includegraphics{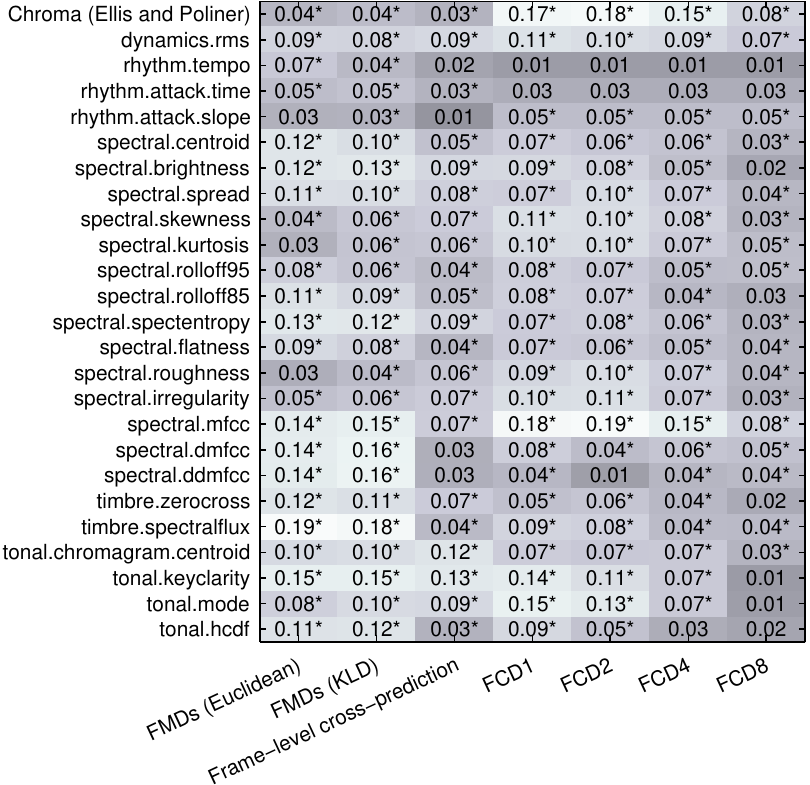}
	\caption{Feature-wise absolute correlation $|\tau_b|$ between pairwise distances and web-sourced similarity annotations. Pairwise distances respectively obtained using FMDs compared using Euclidean distance and KLD (first and second columns), cross-prediction (third column), Euclidean distance applied to FCDs (remaining columns). Starred entries indicate significance, where we apply Bonferroni correction to $\alpha = 0.05$.}
    \label{fig:featurewisecorrelation}
\end{figure}

We examine the correlation between descriptor distances and five-point scale similarity ratings across individual audio features. Fig.~\ref{fig:featurewisecorrelation} depicts correlations $\tau_b$ for FCDs and FMDs, where we compare FMDs using both Euclidean distance and KLD. In addition to FMDs, as described in Section~\ref{sec:distancemeasures} we consider as a baseline the cross-prediction error. 

We observe that FCDs and FMDs both yield maximum correlation 0.19 (comparing FCD2 to FMDs, with both distances computed using Euclidean distance); similarly, FMDs compared using KLD yield maximum correlation 0.18. Across descriptors, with $\alpha=0.05$ and applying Bonferroni correction, the majority of features yield significant correlations. In contrast, for cross-prediction, effect sizes are comparatively small. Comparing descriptors, for FCD2 we observe correlations exceeding 0.1 for 9 features, and for 12 features for the case of FMDs compared either using KLD or Euclidean distance. On average, FMDs yield greater correlation compared to FCD1 (0.095 versus 0.087). However, for specific features FCDs yield higher correlation than FMDs. Comparing FCDs amongst temporal resolutions, we observe a monotonically decreasing relationship between downsampling factor and average correlation.

\begin{figure}[h]
	\begin{minipage}{.99\linewidth}
      \centering
	  \centerline{(a) Five-point rating scale}
      \centerline{\includegraphics{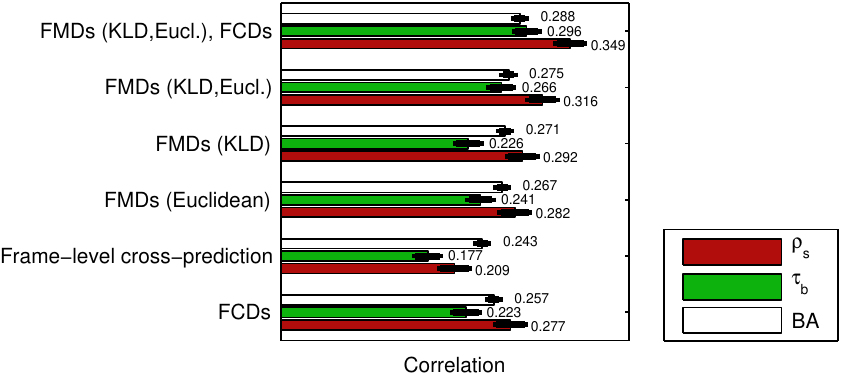}}
	\end{minipage}

	\begin{minipage}{.99\linewidth}
      \centering
	  \centerline{(b) Four-point rating scale}
      \centerline{\includegraphics{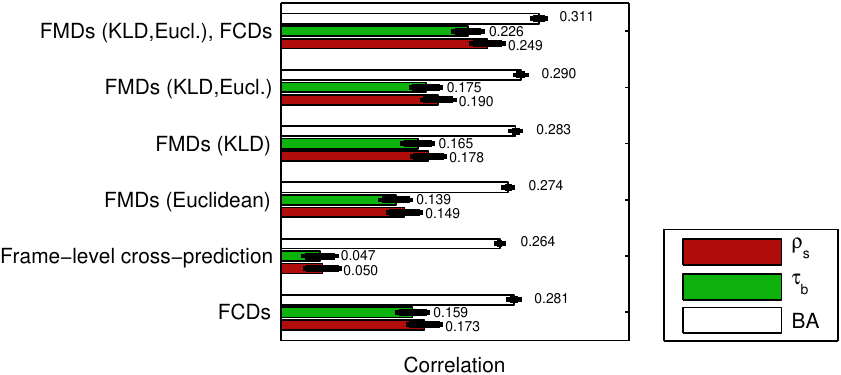}}
	\end{minipage}
	\caption{Similarity rating prediction accuracy. Standard errors obtained by bootstrap sampling pairs of predicted and observed similarity ratings.}
    \label{fig:combinedcorrelation}
\end{figure}

Fig.~\ref{fig:combinedcorrelation} displays a comparison of similarity rating prediction accuracy, where for each descriptor set in Table~\ref{tab:yearprediction} we apply feature selection as described in Section~\ref{sec:modelestimation}. We estimate models using $\tau_b$, $\rho_s$, BA as performance statistics. We consider both 5-point and 4-point rating scales. In particular, we consider the performance gain obtained by including FCDs in our models.

Across both rating scales, we observe that FCDs are outperformed by FMDs compared using KLD alone, or using Euclidean distance and KLD in combination.  However, a combination of FCDs and FMDs outperforms evaluated combinations employing FMDs alone. By incorporating compression descriptors, compared to FMDs based on aggregated KLD and Euclidean distance, based on the five-point rating scale we obtain absolute performance gains of $0.033$, $0.030$, $0.013$ with respect to $\rho_s$, $\tau_b$, BA. The respective relative performance gains are 10.4\%, 11.3\%, 4.7\%. Based on the four-point rating scale, we obtain absolute performance gains of $0.059$, $0.051$, $0.021$; the respective relative performance gains are 31.1\%, 29.1\%, 7.2\%. For the model using $\rho_s$ and the four-point rating scale, Table~\ref{tab:predictionsconfusionmatrix} displays confusion matrices of predicted versus annotated ratings. We test for differences between correlations by applying bootstrap sampling to predicted and observed similarity ratings, from which in turn we estimate standard errors of performance statistics. Based on a one-way analysis of variance with Tukey-Kramer post-hoc analysis and setting $\alpha=0.05$, we reject the hypothesis of no difference between correlations across all considered pairs, for all considered performance statistics.


\begin{figure}[h]
    \centering
    \includegraphics{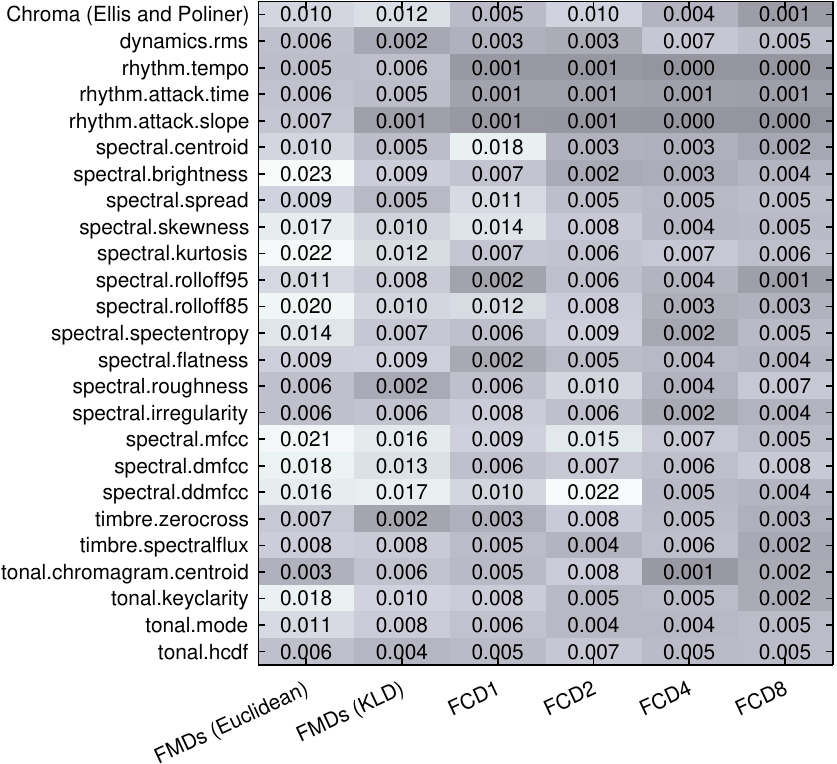}
	\caption{Normalised regression coefficient magnitudes, estimated using elastic net regression, for task of similarity rating prediction. Candidate descriptor set comprised of FCDs compared using Euclidean distance, and FMDs compared using Euclidean distance and KLD.}
    \label{fig:selectedfeatures}
\end{figure} 

\begin{table}
\begin{minipage}{.48\linewidth}
\centering
\centerline{(a) FMDs (KLD, Eucl.)}
\begin{tabular}{c c | c c c c}
 & & \multicolumn{4}{c}{\scriptsize Predicted} \\
 & & \textbf{1;2} & \textbf{3} & \textbf{4} & \textbf{5} \\
\hline
\multirow{5}{*}{\rotatebox{90}{\scriptsize Annotated}}
 & \textbf{1;2} & 1490 & 96 & 80 & 1 \\
 & \textbf{3} & 532 & 75 & 78 & 0 \\
 & \textbf{4} & 435 & 48 & 87 & 4 \\
 & \textbf{5} & 130 & 15 & 42 & 0 
\end{tabular}
\end{minipage}
\begin{minipage}{.48\linewidth}
\centering
\centerline{(b) FMDs (KLD, Eucl.), FCDs}
\begin{tabular}{| c c c c c}
 \multicolumn{4}{c}{\scriptsize Predicted} \\
 \textbf{1;2} & \textbf{3} & \textbf{4} & \textbf{5} \\
\hline
  1361 & 152 & 131 & 23 \\
  458 & 115 & 101 & 11 \\
  311 & 111 & 129 & 23 \\
  106 & 37 & 37 & 7 \\
\end{tabular}
\end{minipage}
\caption{Confusion matrices of predicted versus annotated similarity ratings, for model based on four-point rating scale and $\rho_s$.}
\label{tab:predictionsconfusionmatrix}
\end{table}

Fig.~\ref{fig:selectedfeatures} displays regression coefficients across features and descriptor classes, where we consider the best-performing model evaluated in Fig.~\ref{fig:combinedcorrelation} based on $\rho_s$ and using the five-point rating scale. We sum regression coefficient magnitudes across each of the $K$ binary classifiers given in (\ref{eqn:multinomialregressionmodel}), before normalising the obtained values to sum to one. Comparing FMDs and FCDs, we observe that both FCDs and FMDs are selected within individual features. FCDs appear to be selected across diverse temporal resolutions, with emphasis on higher temporal resolutions. We observe that multiple FCD resolutions are selected within the same feature.

\subsection{Song year prediction}
\label{sec:evaluationsongyearprediction}
\begin{figure}[h]
	\centering
    \includegraphics[scale=0.9]{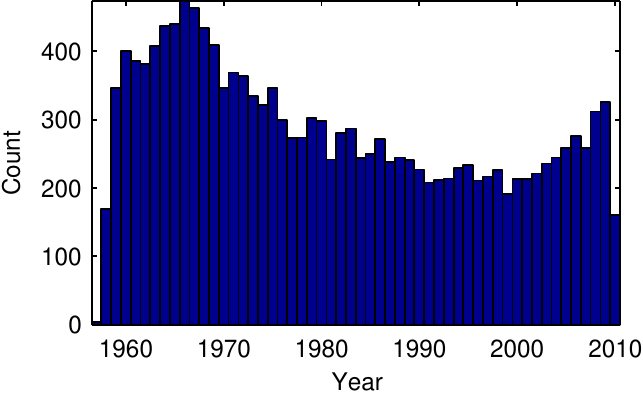}
	\caption{Histogram of chart entry dates.}
    \label{fig:chartentrydatehistogram}
\end{figure}

For song year prediction, we compute FCDs and FMDs as performed for similarity rating prediction. We use chart entry dates as our annotation data and apply the linear regression model given in~(\ref{eqn:regressionmodel}). Fig.~\ref{fig:chartentrydatehistogram} displays a histogram of chart entry dates.

\subsubsection{Model estimation}
\label{sec:modelestimationyearprediction}
To evaluate our descriptors for song year prediction, we partition the dataset into random training and testing subsets, where we ensure that title or artist strings are not duplicated across subsets. We apply the aforementioned filtering procedure to control for potential cover version and album effects, in addition to any analogous effects at the level of artists \cite{flexer2010effects}. The resulting training and testing datasets consist of 10\,728 and 4\,745 tracks respectively. We deem as outliers descriptor values in the training data exceeding 10 standard deviations beyond the 99th percentile. We replace such outliers with imputed values, using the $K$-nearest neighbour algorithm.

We apply linear regression separately to sets of descriptor vectors, as specified in Table~\ref{tab:yearprediction}. We standardise descriptors by subtracting the mean and dividing by the variance of the training data. As performed for similarity rating prediction, we compute FCDs separately across temporal resolutions and across audio features. In contrast, we apply linear regression directly to descriptor vectors without the intermediate step of computing distances. Based on a set of 25 audio features, given a single track we obtain a total of 300 scalar-valued FCDs, for each of which we estimate a single regression coefficient. Note that since we represent FMDs using the mean and standard deviation, we estimate two regression coefficients for each univariate audio feature. For FMDs, it follows that we estimate 24 regression coefficients for MFCCs and chroma features.

As was performed for similarity rating prediction, we estimate linear regression parameters using ENR. We denote with $\greekvec{\theta} = (\greekvec{\theta}_1^T, \ldots, \greekvec{\theta}_N^T)^T$, $\alpha$ regression coefficients and the model intercept as given in (\ref{eqn:regressionmodel}). Using ENR, we solve
\begin{equation}
  \min_{\greekvec{\theta}, \alpha} \left\{ \eta \left( \nu \|\greekvec{\theta} \|_1 +  \left(1 - \nu \right) \frac{1}{2} \| \greekvec{\theta} \|_2^2 \right) + \mathrm{SSR}(\greekvec{\theta}, \alpha) \right\}
\end{equation}
where $\mathrm{SSR}(\greekvec{\theta}, \alpha)$ denotes the sum of squared residuals. Both $\eta$, $\nu$ behave as defined in (\ref{eqn:enrmultinomialregression}). We apply cross-validation to training data and optimise $\eta$ by determining minimal prediction mean square error. We again consider $\nu$ a hyper-parameter which we assign constant value; we optimise prediction mean square error based on a model incorporating FCDs and FMDs, and by applying cross-validation to training data. We threshold predictions to fall in the range $[1957\text{y}\,..\,2010\text{y}]$.

In addition to the year prediction task based on individual tracks, we evaluate prediction performance when considering groups of tracks. We perform this experiment to establish whether FCDs consistently improve performance when combined with grouped FMDs, or if grouped FMDs amortise any potential performance gain due to FCDs. We select groups of tracks by applying a non-overlapping sliding window to chart entry dates. We then take as descriptor vector $\vec{r}'_{w,n}$ the average
\begin{equation}
  \vec{r}'_{w,n} = \frac{1}{|C_w|} \sum_{i \in C_w} \vec{r}_{i,n}
\end{equation}
where $C_w$ denotes the set of tracks at window position $w$. We apply the windowing procedure separately to training and testing data sets. Note that by windowing tracks, at each window position we assume prior knowledge of differences among chart entry times in training and testing data, respectively. For a given window size, we average descriptor vectors in the training data and proceed as described in Section~\ref{sec:modelestimationyearprediction}. Given the obtained regression model and given averaged descriptor vectors at window position $z$ in the testing data, we seek to predict the associated window centre $y'_z$.

\subsubsection{Performance statistics}
We quantify prediction accuracy with respect to annotated chart entry dates, using the mean absolute error (MAE) and root mean square error (RMSE) statistics.

\subsubsection{Results}
\begin{figure}[th]
  \begin{minipage}{.49\linewidth}
    \centering
	\centerline{
    	\includegraphics[angle=90]{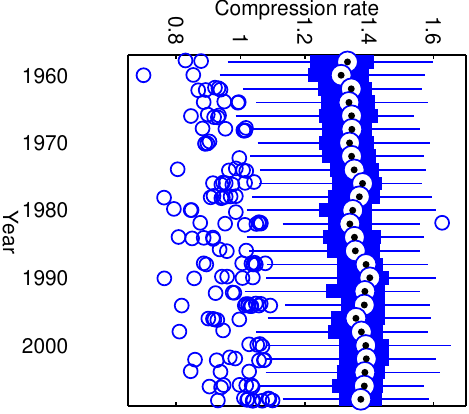}
	}
    \centerline{(a) FCDs (FCD1)}\medskip
  \end{minipage}
  \begin{minipage}{.49\linewidth}
    \centering
	\centerline{
    	\includegraphics[angle=90]{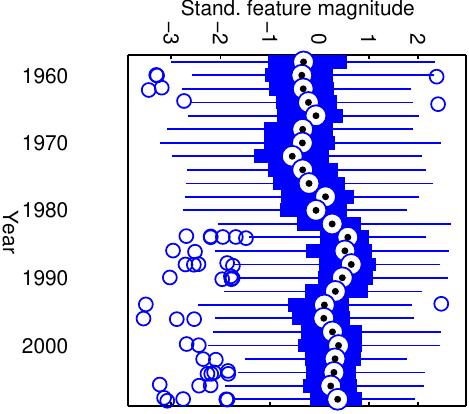}
	}
    \centerline{(b) FMDs}\medskip
  \end{minipage}
	\caption{Box plots of FCDs and FMDs computed using spectral spread features, with FCDs computed without downsampling. Each box corresponds to the position of a non-overlapping 1-year window applied to chart entry dates.}
    \label{fig:boxplotfeatureovertime}
\end{figure}

\begin{figure}[h]
  \begin{minipage}{.49\linewidth}
    \centering
    \centerline{
    	\includegraphics[scale=.95]{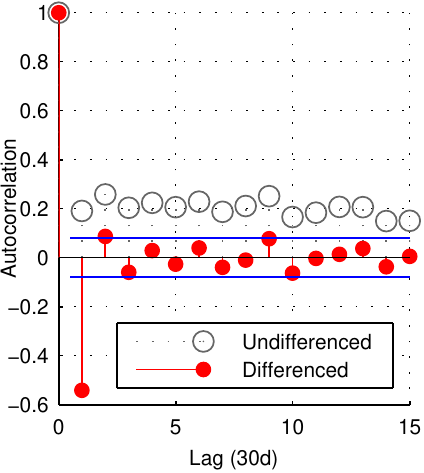}
	}
    \centerline{(a) FCDs (FCD1)}\medskip
  \end{minipage}
  \begin{minipage}{.50\linewidth}
    \centering
    \centerline{
		\includegraphics[scale=.95]{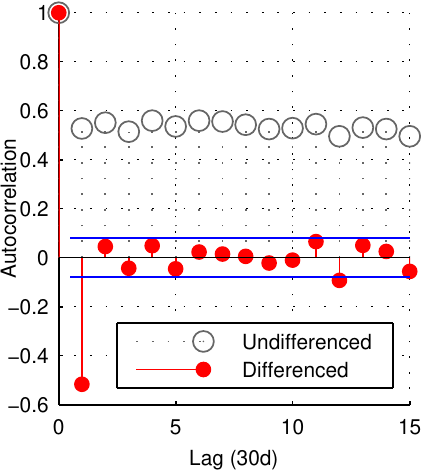}
	}
    \centerline{(b) FMDs}\medskip
  \end{minipage}
	\caption{Sample autocorrelation of undifferenced and differenced FCD, FMD averages. Descriptor averages obtained by applying non-overlapping 30-day window to chart entry dates. Descriptors computed on spectral spread features, with FCDs computed without downsampling. Horizontal bars indicate 95\% confidence intervals under the assumption of Gaussian white noise for differenced time series.}
    \label{fig:acf}
\end{figure}

Fig.~\ref{fig:boxplotfeatureovertime} displays the result of exploratory analysis for song year prediction, where for FMDs and FCDs we group descriptor values across time, by applying a non-overlapping 2-year sliding window to chart entry dates. We restrict analysis to obtained spectral spread features \cite{lartillot2007matlab}. The resulting year-wise box plots suggest that the examined descriptors are non-stationary with respect to chart entry dates, exhibiting distinct trends. To examine the behaviour of descriptors at a finer time scale, we apply a non-overlapping 30-day sliding window to chart entry dates, where at each window position we compute the mean descriptor value. Examining the sample autocorrelation of the resulting time series for lags in the range $[1\,..\,15]$, we observe weaker correlations for FCDs compared to FMDs. Yet, both autocorrelations exhibit slowly decaying autocorrelations (Fig.~\ref{fig:acf}), characteristic of a non-stationary time series \cite{kirchgassner2012introduction}. Following the method of Box and Jenkins \cite{box2013time}, we attempt to attain stationarity by applying first-order differencing to the time series. However, we observe autocorrelation close to $-0.5$ at unit lag, suggesting that the time series have been overdifferenced \cite{kirchgassner2012introduction}. We interpret these observations as evidence for a non-trivial, trend-exhibiting process governing observed descriptor values \cite{granger1980introduction}. 

\begin{table}[h]
\centering
\begin{tabular}{l c c}
\hline
\hline
Set & MAE & RMSE \\
\hline
FCDs & 9.44 $\pm$ 0.096 & 11.54 $\pm$ 0.107 \\
FMDs & 8.28 $\pm$ 0.092 & 10.45 $\pm$ 0.113 \\
Combined & 7.38 $\pm$ 0.085 & 9.43 $\pm$ 0.107 \\
\end{tabular}
\caption{Summary of song year prediction accuracy, expressed using MAE and RMSE statistics. Standard errors obtained by bootstrap sampling pairs of predicted and observed chart entry dates.}
\label{tab:yearpredictionaccuracy}
\end{table}

Table~\ref{tab:yearpredictionaccuracy} summarises the accuracy of song year prediction using MAE and RMSE statistics. Quantified using either MAE or RMSE, song year prediction based on FMDs outperforms prediction using FCDs alone. However, we observe that a combination of FMDs and FCDs yields the highest prediction accuracy. By incorporating FCDs we observe performance gains of 10.9\%, 9.8\% relative to FMDs, in terms of MAE and RMSE. As performed in Section~\ref{sec:similarityratingpredictionresults}, we test for differences among prediction accuracies by applying bootstrap sampling to predicted and observed chart entry times, from which we estimate standard errors of MAE and RMSE statistics. Again using one-way analysis of variance with Tukey-Kramer post-hoc analysis and setting $\alpha=0.05$, we reject the hypothesis of no difference between prediction accuracies across all pairs, for both MAE and RMSE.

\begin{figure}[h]
    \centering
    \includegraphics{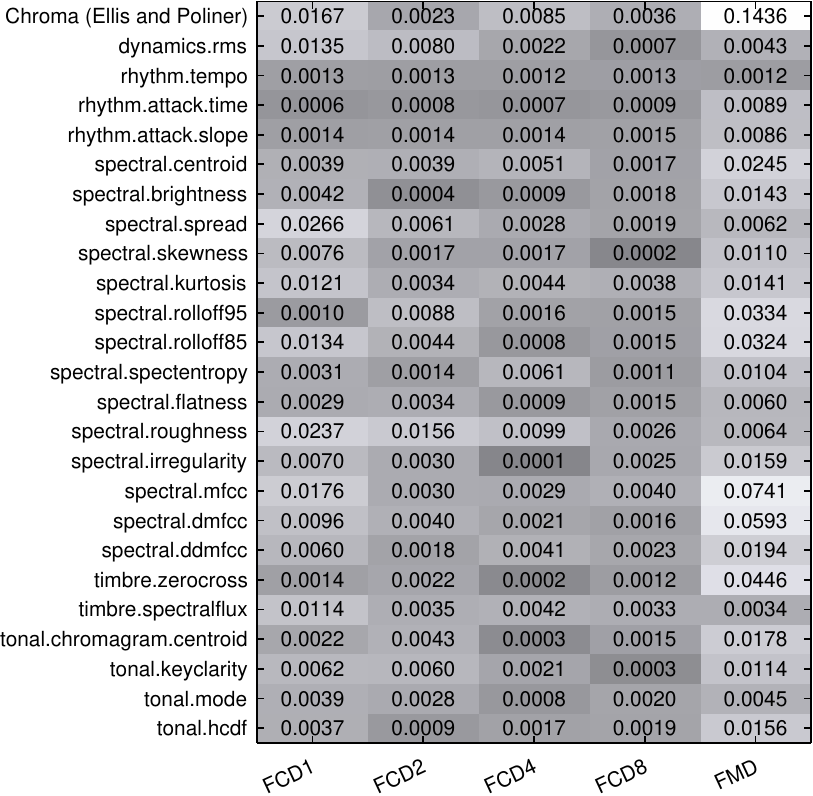}
	\caption{Normalised regression coefficient magnitudes, estimated using elastic net regularisation, for task of song year prediction. Candidate descriptor set comprised of FCDs and FMDs.}
    \label{fig:predictioncoefficients}
\end{figure}

Fig.~\ref{fig:predictioncoefficients} displays regression coefficients obtained using unwindowed chart entry dates. We compute coefficient magnitudes and normalise to sum to one. Thus computed, we interpret coefficient magnitudes as predictive utilities across individual audio features. In addition, we consider the utility of FCDs across time scales, compared to FMDs. Summed across features, we observe that compared to FCD1, FMDs are weighted more strongly (0.591 versus 0.201). Further examining relative weightings, we observe a prevalence of weight assigned to FCD1 compared to higher downsampling factors. However, we observe that individual features may be weighted relatively strongly across multiple temporal scales. Note from Table~\ref{tab:similarityprediction} that for chroma features, MFCCs and derivatives, FMD weights are summed across 24 prediction coefficients, compared to 3 coefficients for FCDs.

\begin{figure}[h]
    \centering
    \includegraphics[scale=0.8]{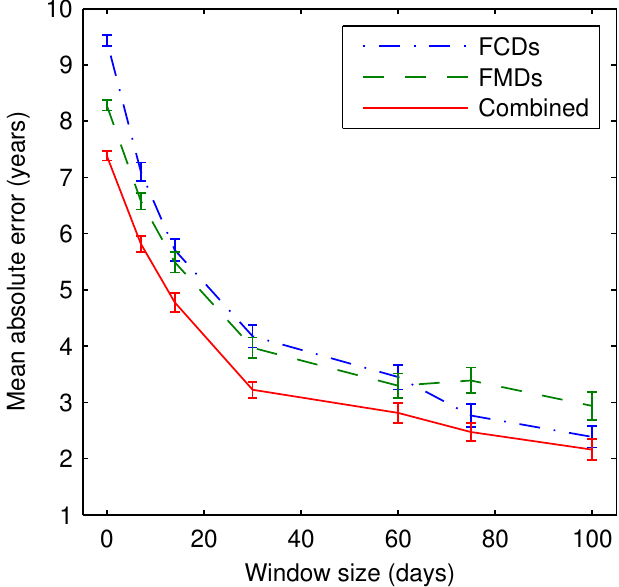}
	\caption{Song year prediction accuracy obtained using windowed descriptors, in response to window size. Error bars denote standard errors.}
    \label{fig:yearpredictionperformance}
\end{figure}

In Fig.~\ref{fig:yearpredictionperformance} we examine prediction accuracy in response to windowed descriptors, as described in Section~\ref{sec:modelestimation} and quantified using MAE. For increasing window size up to 60d, performance improves monotonically across all considered descriptor sets. Across considered window sizes, using combined FCDs and FMDs we observe a mean performance gain of 17.5\%, relative to using FMDs alone.

\section{Conclusions}
\label{sec:conclusions}
We have considered the problem of determining musical similarity, using feature sequences extracted from musical audio. In particular, we have considered musical similarity in the context of two low-specificity content retrieval tasks, namely similarity rating prediction and song year prediction. To this end, we have evaluated the utility of sequential complexity as a descriptor for quantifying musical similarity.

For both considered tasks, we observe that sequential complexity descriptors predict the outcome variable. Furthermore, in combination with feature moment descriptors, sequential complexity descriptors improve prediction accuracy with respect to the baseline. The results confirm that our proposed descriptors capture musically relevant information and that temporal structure is relevant in our chosen domain. Consequently, our results show that sequential complexity may be used to improve the accuracy of low-specificity content retrieval based on bag-of features approaches.

Our proposed descriptors are computed in an unsupervised manner and may be implemented efficiently, requiring $O(n)$ time complexity for each track \cite{effros2000ppm}. In addition, our proposed descriptors have similar dimensionality compared to feature moment descriptors. Since our descriptors may be computed off-line or incrementally and thereafter combined with indexing methods as proposed in \cite{slaney2008locality,rhodes2010investigating,schluter2013}, we deem them potentially applicable in large-scale content retrieval systems.

Similar to results obtained in \cite{foucard2011multi,dielemanmultiscale,hamel2012building,hamel2011temporal}, our results using sequential complexity descriptors suggest that an approach based on multiple temporal resolutions is advantageous for determining musical similarity. As an alternative to downsampled features, we initially employed beat-synchronous representations, which yielded comparatively small gains in prediction accuracy, when combined with original frame-based features. This result suggests that for our chosen domain, temporal structure at short time scales is more advantageous, compared to temporal structure at the metrical level. One possible explanation for this behaviour is that an abundance of observations is beneficial when estimating compression rates. Alternatively, for our chosen tasks similarity judgements might predominantly be based on short-term timbral characteristics, rather than long-term structures such as motifs and chord progressions. For future work, we aim to examine in closer detail the utility of representing features at multiple time scales, and to characterise the feature spaces relevant for similarity judgements.

For similarity rating prediction, note that by biasing towards tracks with proximate chart entry dates, we attempt to control for historical changes in audio production. For song year prediction, where we do not control in the described manner, audio production may confound the association between musical similarity and chart entry date. We acknowledge that in both cases, audio production may confound the association between similarity measures and respective outcome variables, as observed in \cite{sturm2012two}. For future work, we aim to measure the degree of confounding by introducing suitable audio degradations \cite{mauch2013audio}. A further issue concerns the practical impact of predicted similarity in music information retrieval. We aim to evaluate our descriptors for search, navigation and recommendation tasks, using collections of various scales.

Finally, the present work considers only a single sequential complexity measure, estimated using a single algorithm. It is conceivable that using multiple compression algorithms may reduce the error variance of estimated sequential complexity. Using alternative classification tasks, we aim to evaluate whether multiple compressors yield an improvement in prediction accuracy.

\section{Acknowledgements}
This work benefited from advice and comments from Andrew J.~R.~Simpson, Dan Stowell, Anssi Klapuri, Mark D.~Plumbley, and Armand Leroi.

\opt{journal}{
  \ifCLASSOPTIONcaptionsoff
    \newpage
  \fi
}

\bibliographystyle{IEEEtran}
\bibliography{refs}

  \vspace*{-2\baselineskip}
\end{document}